\documentclass[aip,graphicx,reprint]{revtex4-1}

\usepackage{booktabs} % for much better looking tables
\usepackage{array} % for better arrays (eg matrices) in maths
\usepackage{paralist} % very flexible & customisable lists (eg. enumerate/itemize, etc.)
\usepackage{verbatim} % adds environment for commenting out blocks of text & for better verbatim
\usepackage{amsmath} %for more math symbols
\usepackage{amssymb} %for even more math symbols
\usepackage{url} %for hyperlinks
\usepackage{epstopdf} %allows importing eps figures for pdfLatex
\usepackage{color}
\usepackage{stackengine}
\usepackage{graphicx}
\usepackage{hyperref}
\usepackage{subcaption}

\draft % marks overfull lines with a black rule on the right

\begin{document}

%\preprint{}

\title{Temporal Network Epistemology:\\on Reaching Consensus in Real World Setting} %Title of paper



\author{Rados{\l}aw Michalski}
\email[]{E-mail: radoslaw.michalski@pwr.edu.pl}
\thanks{\\The following article has been submitted to Chaos. After it is published, it will be found at \href{https://aip.scitation.org/journal/cha}{the journal's webpage}.}
%\thanks{}
\author{Damian Serwata}
\author{Mateusz Nurek}
\affiliation{Department of Artificial Intelligence, Faculty of Information and Communication Technology, Wroc{\l}aw University of Science and Technology, Poland}

\author{Boleslaw K. Szymanski}
\affiliation{Department of Computer Science, Rensselaer Polytechnic Institute, Troy, NY, USA}
\affiliation{Spo{\l}eczna Akademia Nauk, {\L}{\'o}d{\'z}, Poland}

\author{Przemys{\l}aw Kazienko}
\affiliation{Department of Artificial Intelligence, Faculty of Information and Communication Technology, Wroc{\l}aw University of Science and Technology, Poland}

\author{Tao Jia}
\affiliation{Southwest University, Chongqing, China}


\date{\today}

\begin{abstract}
This work develops the concept of temporal network epistemology model enabling the simulation of the learning process in dynamic networks. The results of the research, conducted on the temporal social network generated using the CogSNet model and on the static topologies as a reference, indicate a significant influence of the network temporal dynamics on the outcome and flow of the learning process. It has been shown that not only the dynamics of reaching consensus is different compared to baseline models but also that previously unobserved phenomena appear, such as uninformed agents or different consensus states for disconnected components. It has been also observed that sometimes only the change of the network structure can contribute to reaching consensus. The introduced approach and the experimental results can be used to better understand the way how human communities collectively solve both complex problems at the scientific level and to inquire into the correctness of less complex but common and equally important beliefs' spreading across entire societies.
\end{abstract}

\pacs{}% insert suggested PACS numbers in braces on next line

\maketitle %\maketitle must follow title, authors, abstract and \pacs

\begin{quotation}
The human learning process would not be possible without social relationships that allow communities to exchange views and to learn through social interactions. The analysis of how this process works led to the development of a network epistemology field that studies how people exchange information and make decisions based on acquired knowledge. Previous research in this area has focused on static networks which do not capture the dynamics of connectivity changes that are present in real social networks. In this work, we analyzed the impact of network dynamics on social learning and proposed a new model closer to the real-world scenario.
\end{quotation}

\section{Introduction}
Human beings are extremely social. The specific abilities that our species has developed in the course of the cognitive revolution have allowed us to dominate other organisms on the planet. One crucial component in this process is the incredibly powerful and effective, evolved ability of homo sapiens to communicate with one another, exchange and store large amounts of information. The fundamental role of this process is evident in jumps of civilization sophistication when storage of information moved,  thanks to the invention of the alphabet, from individuals memorizing cultural and religious myths, to storing information on stones, papyrus, or paper. This made access to information easier but still limited. Not until the invention of print press  democratization of access had been possible thanks to massive and inexpensive information replication technology. The final step, the invention of the internet and digital storage brought the cost of access to information nearly to nil and started the information revolution. Initially, a few people devoted their lives to memorizing history or replicating manuscripts, so social learning was a necessity in times of limited access. Today, it is the opposite problem of abundance of information that makes social learning necessary for filtering vast information to establish a consensus and communicate it to the masses. These behaviors have been captured in social learning and social choice theory. Given the evolutionary history of our species, it should come as no surprise that human systems for forming views and opinions are also social in nature. Humans pass information from one person to another, and as a result, a large number of our views are dependent on the environment of our contacts. This inherent human way of spreading knowledge is the foundation of today's extremely complex and advanced culture and science. However, the social learning process has drawbacks that sometimes make reaching consensus impossible. Research has shown that people often use heuristic thinking, which on the one hand preserves their energy but on the other hand results in confirmation bias~\cite{sherman1984cognitive, goldstein2002models, knobloch2020confirmation}. Individuals will seek confirmation of their opinion to the point that they will stick to it even if it is false. Furthermore, they will interact more often with people who hold a similar opinion. Flamino et al.~\cite{flamino2021creation} demonstrated that groups strive to optimize the utility of interaction of their members, which results in a total polarization of opinions within the groups. An interesting fact about human social learning is revealed by research comparing cognition between humans and other primates~\cite{hopper2012primate, tomasello2010ape, hopper2008observational}. Research shows that social learning and trusting authority among humans is much stronger than in apes~\cite{horner2005causal, seed2011causal}. \par
% Modeling
Due to the aforementioned role of the social aspect of information exchange in the social learning process~\cite{zollman2007communication}, researchers for long have been carrying out research that helps to understand the characteristics of this process and its complexity. There is a growing interest in modeling in the scientific community. What is the best community structure for spreading information? What conditions must be met to allow the community to reach a consensus on a given topic? What influences the speed of the spread of information and opinions? Various models developed by researchers in recent years provide answers to these questions. Through the use of mathematical modeling, it is possible, to a certain extent, to predict spreading scenarios and to gain insight into the complex dynamics of the social learning process. This research is extremely important and useful, but often the experiments are conducted on static and synthetic network structures. Therefore, the results of such research should always be approached with caution and, if possible, verified. \par
% Network Epistemology Model
Many of these models are based on the network structure, which has been identified as a major factor in determining how societies move towards adopting ideologies. The network epistemology framework was initially proposed in~\cite{bala_goyal} to model learning from neighbors. Subsequently, using that framework, Zollman~\cite{zollman2007communication} created a network epistemology model that provides an interesting way of capturing aspects of social learning in human societies and its impact on theory adoption. It is an agent-based model that offers an imitation of a human and especially a scientific way of analyzing information and updating beliefs. It has so far mainly been used to model the spread of knowledge in scientific communities, but it can also be applied in a more general context, i.e. to study any community of people receiving evidence and sharing beliefs. The authors of~\cite{oconnor_weatherall_2018_3, oconnor_weatherall_2018} used agent-based modeling to study the effects of polarization and conformity on reaching consensus. They showed that through social trust and conformism, the learning process could be disrupted, which will result in the agent accepting the opinion prevailing in a given group, even if it is not true. In~\cite{oconnor_weatherall_2018_2} they presented a model for exploring belief sharing. They explain the phenomenon that individuals share highly correlated views on many different topics within a community. Finally, the impact of spreading false information in an epistemic network was investigated in~\cite{oconnor_weatherall_2020_2}, and resulted in the conclusion that even a small percentage of propagandists can completely sabotage the learning process in a community. 

Agent-based models of reaching consensus also include a group of simpler binary agreement models that are based on the naming game~\cite{baronchelli2006sharp, dall2006nonequilibrium, lu2008naming, baronchelli2011role} or voter models~\cite{castellano2009statistical, miguel2005binary, vazquez2003constrained}. In the previously mentioned models, each agent's belief was represented on a continuous scale from 0 to 1. In binary agreement models, however, agents are assigned one or both of the competing viewpoints. The conducted research shows that this model is able to guarantee consensus for individuals organized in a complete graph structure and the time in which this happens depends on the number of agents in the network~\cite{castello_baronchelli_vittorio_2009}. Xie et al.~\cite{szymanski_2011} show that even a small fraction of agents resistant to social influence and constantly proclaiming their fixed views can be sufficient to convince an overwhelmingly large proportion of the community to adopt a new view.  Furthermore, in~\cite{kearns_2009}, the authors analyze an interesting property of the binary consensus model, where individuals promoting their own views seek a broader consensus. Also, neutral or uninformed agents can prevent consensus to any of the committed groups~\cite{thompson2014propensity} even in groups of animals~\cite{couzin2011uninformed}.

In the field of opinion dynamics, the continuous Deffuant model~\cite{deffuant2000mixing} is often used to simulate interactions between individuals and the spread of beliefs. A random pair of nodes is sequentially selected to update their opinion if they share similar beliefs (the difference in opinion must be below a given confidence bound). The Hegselmann-Krause model~\cite{hegselmann2002opinion} updates the opinion of all nodes simultaneously taking into account the opinion of neighbors, which can reflect the influence of a group on an individual in the real world.

The study of consensus and collective behavior in multi agent systems is not limited to the problem of social learning but also has applications in engineering, physics, robotics, and biology. Consensus dynamics is an important issue of cooperative control in the area of synchronization of coupled oscillators~\cite{li2009consensus, barrat2008dynamical}, multi-vehicle systems~\cite{ren2007multi}, robotic and biological swarms~\cite{chamanbaz2017swarm, cavagna2018physics}. Decentralized consensus has also found applications in the area of social networks~\cite{fowler2010cooperative, amelkin2017polar, de2022consensus}.

A non-agent-based approach is used to analyze the temporal patterns of citation networks to determine scientific consensus~\cite{shwed2010temporal, light2015scientific}. Using modularity as a measure of consensus over dynamic time windows uncovers consensus evolution. The emergence of consensus creates one common core community with multiple micro-communities in the network, but it can also split the network into two or more competing communities.
\par
% Temporal networks
An important feature of the social networks studied by the researchers is their temporality -- acquaintances and friendships arise and dissolve, old relationships may decay when new people are met~\cite{holme2012temporal}. Most real networks are temporal and their structure -- the configuration of nodes and edges -- changes over time. The temporal structure of link activity can influence the dynamics of interaction in a network just like its topology~\cite{masuda2013temporal}. Researchers have already proposed many models, which attempt to capture this aspect of social networks in different ways~\cite{holme2015modern}. Recently introduced cognition-driven social network (CogSNet) model~\cite{michalski2021social} presents a unique approach based on social perception and due to incorporating cognitive processes on forgetting it is used in this work as a foundation of a temporal social network. \par

The previously mentioned methods of agent-based consensus modeling were developed for static graphs, abstracting from  temporal characteristics of many real networks and limiting the consensus dynamics. However, in the last decade, an adaptation of the binary voter model~\cite{hoppe2013mutual, fernandez2013timing} has been attempted, and a continuous model has been proposed for activity-driven networks~\cite{li2018opinion}. A common feature of the above works is their focus on synthetic networks. In recent years, Li et al.~\cite{li2019impact} analyzed the effect of temporal networks on reaching consensus. The authors, based on the Deffuant model and real social networks, showed that reaching consensus in temporal networks takes more time than in static networks. The problem of consensus has also been studied in temporal hypergraphs~\cite{neuhauser2021consensus}.

% Content
This work addresses the topic of social learning modeling in temporal social networks. The question is whether the currently used models correctly represent the dynamics of opinion propagation in real social networks. Special attention is given to the network epistemology model, for which so far no extension has been developed to make the model work for temporal networks, and we advocate here for such an extension the use of the CogSNet temporal network representation model. Unlike previous work, our approach to modeling consensus based on an epistemic framework uses not only beliefs but also evidence collected by agents. We test our approach on a real temporal network that contains interactions over a significant period of 110 days (cf. Li et al.~\cite{li2019impact} that uses up to one week periods). Moreover, our continuous temporal network model is based on a stream of events, which eliminates the problem that the size of the time window affects the network structure. An additional advantage of CogSNet is that it models the cognitive abilities of the human brain.

\par

\section{Materials and Methods}
\subsection{Epistemic Model}
Epistemic models are tackling the problem of collaborative learning that leads to developing new ideas and beliefs that are hopefully shared by all members of a group. As presented in~\cite{stahl2004building}, collaborative learning can be understood as a particular way in which a group may construct a new degree of understanding about the topic that they are investigating. However, this cannot be done individually by each member, but human interactions are required to facilitate consensus reaching~\cite{gavsevic2019sens}. The substantial base and foundation for the later versions of the network epistemology model were laid by Bala and Goyal in~\cite{bala1998learning}. In this work we employ a specific version of this model, that was brought to the field of epistemology by Zollman in \cite{zollman2007communication}. The basic concept assumes that the model consists of a fixed set of \textit{K} agents that represent a group of people. The agents are connected to each other by symmetric relationships in a graph structure. These agents are faced with the problem of choosing between two theories with different effectiveness defined as $\textit{A} = \{\textit{Alpha}, \textit{Beta}\}$. In this specific case, the dilemma can be described as a two-armed bandit problem with two possible states of the world $\Theta = \{\theta_0, \theta_1\}$. The \textit{Alpha} arm returns a payoff equal to 1 with a probability $p_{Alpha} = 0.5$ in both states. The \textit{Beta} arm returns the same payoff but with a different probability equal to $p_{Beta} = 0.5 + \epsilon$, where $|\epsilon| < 0.5$ and $\epsilon$ takes positive values in $\theta_0$ and negative values in $\theta_1$. All agents know the effectiveness of the \textit{Alpha} arm, but they do not know whether the \textit{Beta} arm is better ($\theta_0$) or worse ($\theta_1$). The problem comes down to choosing the better arm. As an example, a group of doctors may use two different therapies with different efficacy to treat a certain disease. One of these treatments has been used for a long time and its efficacy is well known, and the other, innovative one has similar but unknown efficacy that may be better or worse.

The process of belief evolution in this model is carried out in an iterative manner, with iterations indexed by $i = 1, 2, ...$. The belief of agent \textit{a} in iteration \textit{i} is represented by the credence $c_{a,i} \in [0,1]$. This value can be interpreted as the probability that the \textit{Beta} action is better than the \textit{Alpha} action. The action $A_{a,i}$ chosen by an agent \textit{a} in iteration \textit{i} is determined as follows:
\begin{equation}
A_{a,i} = 
\begin{cases}
Alpha & \text{if } c_{a,i} \leq 0.5 \\
Beta & \text{if } c_{a,i} > 0.5
\end{cases}
\end{equation}
According to it, if an agent's credence is greater than 0.5, the \textit{Beta} action is chosen, otherwise -- the \textit{Alpha} action is performed. 

In the beginning, the agents' credence is drawn from a uniform distribution over the interval of possible values. Then, in each iteration, agents take two activities: experimenting and updating beliefs. Within the experimentation phase, each agent performs its preferred action a specified number of times \textit{N} and observes the results. The results of an action are drawn from a binomial distribution with an action-specific probability of success. For example, an agent whose credence is equal to 0.8 has 80\% confidence that the action \textit{Beta} is better than \textit{Alpha}. Hence, they perform it 10 times, out of which e.g. 6 are successful. Next, each agent updates their credence based on the collected results. Agents do this using Bayes' rule (strict conditionalization). The results of performing the \textit{Alpha} action are not used to update views, as its payoff is known and there is general agreement on its value. The observations are shared among the agents over their network social relationships. The credence update is performed not only on the basis of the evidence observed by the specific agent but also upon the evidence collected by its neighbors. It means that any agent whose neighbors have undertaken a \textit{Beta} action will update their views. 

The credence update for a particular agent \textit{a} in iteration \textit{i} is based on $n_{a,i}$ - the total number of attempts to take the action \textit{Beta} made by \textit{a} and their neighbors and $k_{a,i}$ - the total number of successes observed among the performed attempts. According to Bayesian inference, the formula for calculating the posterior credence is given by
\begin{equation}
c_{a,i+1} = \frac{(0.5 + \epsilon)^{2 k_{a,i} - n_{a,i}} c_{a,i}}{(0.5 + \epsilon)^{2 k_{a,i} - n_{a,i}} c_{a,i} + (0.5 - \epsilon)^{2 k_{a,i} - n_{a,i}} (1 - c_{a,i})}
\end{equation}
%which can be transformed to 
%\begin{equation}
%c_{a,i+1} = \frac{1}{1 + \frac{(1 - c_{a,i}) (\frac{0.5 - \epsilon}{0.5 + \epsilon})^{2 k_{a,i} - n_{a,i}}}{c_{a,i}}}
%\end{equation}

This process runs until the entire community converges to one of the beliefs, reaching consensus or up to a given maximum number of iterations. The consensus occurs when all agents have sufficiently low credence, less than or equal to 0.5 (incorrect consensus), or sufficiently high credence -- above an arbitrary threshold, usually set to 0.99 (correct consensus). In this version of the model, the community usually converges to one of the beliefs in a finite time. \par

\subsection{Modification for Temporal Social Networks}
The network epistemology model described above does not take into account the time factor. We hypothesize that the temporality of the structure is important in the context of the spread of beliefs. Therefore, the concept of such a modification is proposed here. The temporal version is based on the application of the temporal social network model -- CogSNet -- that has been introduced in~\cite{michalski2021social}. The CogSNet model accounts for cognitive aspects of social perception. The model explicitly represents each social interaction as a trace in human memory with its corresponding dynamics. It allows for computing the network state at any point in time -- due to this property, it is used in this study. By applying the CogSNet model, we create a setting in which for each iteration of the decision-making process a new snapshot of a temporal network is being created that is based on human past interactions. With this approach, we operate on a temporal network by exploiting a sequence of consecutive snapshots of the network, where each one of them represents this network at a specific point in time of its lifespan. The dynamic weights of the network relationship correspond to the changing agent's ability (trust or influence level) to adapt beliefs acquired by the neighboring nodes. 

\begin{figure}
    \centering\includegraphics[width=0.45\textwidth]{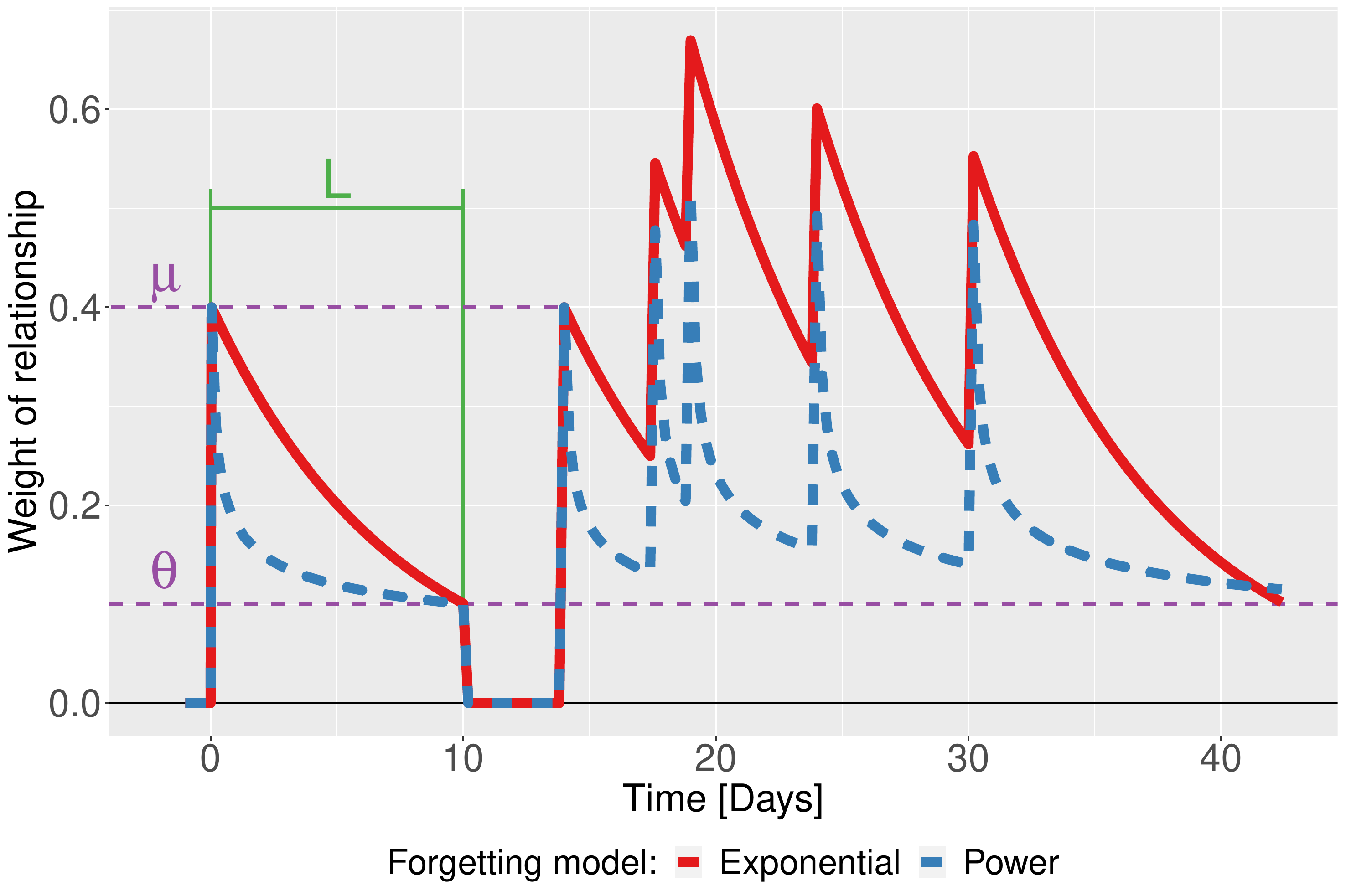}
    \caption{Dynamics of relations in a CogSNet network with exponential and power forgetting functions for parameters $\mu=0.4$, $\theta=0.1$, and $L=10$ days.}
    \label{fig:cogsnet}
\end{figure}

The CogSNet model incorporates a forgetting function~\cite{wixted1991form,wixted1997genuine} and three parameters: $\mu$ indicating the peak that sets the weight when the interaction occurs, $\theta$ being a cut-off threshold, and $\lambda$ that is a parameter for a forgetting function -- the basic idea for the model is presented in Figure~\ref{fig:cogsnet}. These three parameters can be also combined into a trace lifetime $L$ that indicates how long an edge between nodes will live before vanishing if no more interactions happen. Based on previous research, we decided to use the exponential forgetting function and the parameters that most adequately mapped human forgetting, see Table~\ref{tab:cogsnet_params}. More details on how the CogSNet model works are presented in~\cite{michalski2021social}. \par
% Changes
A major change resulting from this temporal approach when compared to static networks is that simulations can only be conducted up to a certain number of iterations, equal to the product of the number of temporal social network states and the number of iterations performed in each state of the network. Therefore, there may be a situation where no consensus is reached by that time. \par
% Isolated nodes
This approach also makes it necessary to introduce some modifications regarding the details of the model. In the case of temporal networks, nodes can lose connection with all their neighbors and be isolated from the rest of the network. Such nodes are labeled as inactive. In contrast, active nodes are those that have at least one neighbor. For this reason, inactive nodes are not considered when checking the consensus condition. Such nodes also do not conduct experiments but retain their credence level and when they return to the network, their belief is the same as it was when they disappeared. \par

\begin{figure*}
    \centering\includegraphics[width=\textwidth]{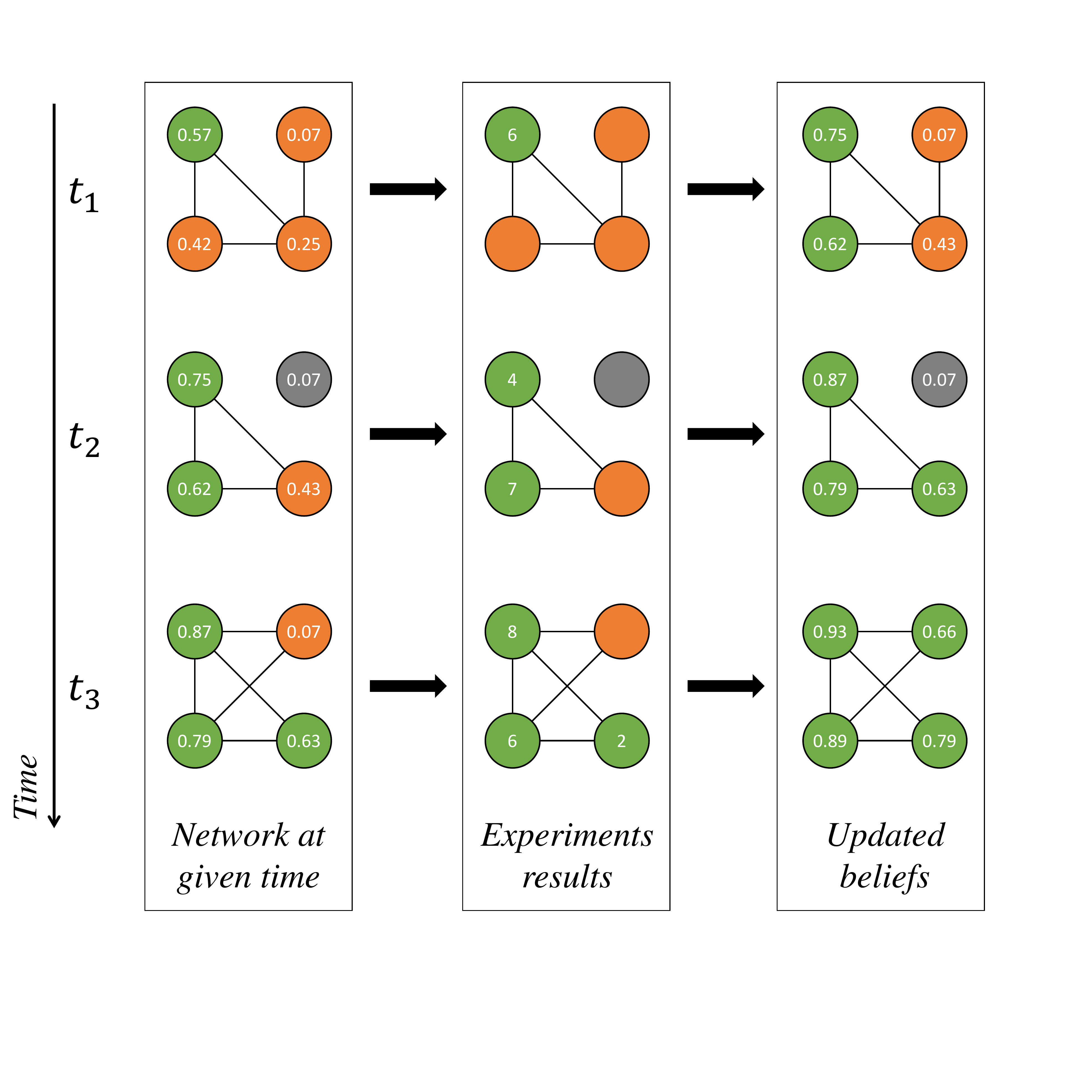}
    \caption{An example of a learning process performed for a temporal network. Rows represent times points \textit{t\textsubscript{1}}, \textit{t\textsubscript{2}} and \textit{t\textsubscript{3}} at which the network is inspected. Columns depict different phases of the learning process. The values in the circles representing nodes for the first and third column mean the credence levels of agents; for the second column - the number of successes returned by Beta action out of 10 tries. Green nodes correspond to agents choosing the Beta action, orange nodes to agents carrying out the Alpha action, and gray ones to disabled agents. For each time point, the network structure is calculated using the CogSNet method and one iteration of the learning process performed on that topology.} 
    \label{fig:modelExample}
\end{figure*}
% Example description
Figure~\ref{fig:modelExample} shows an example of the learning process for a small community. The rows represent states of the temporal network on the timeline and the columns represent phases of the learning process. The credence values ranging from 0 to 1 are assigned to each agent defining their beliefs. The nodes colored in green choose the Beta action which in fact is a better one, and the orange ones choose the worse action Alpha. In this example, the community performs only one learning iteration for one network state. In this iteration, first, the agents that choose the Beta action perform experiments 10 times and count the number of successes to assess if the Beta action is better or worse. Agents opting for Alpha action perform no experiments as the payoff of this action is known and equal to 0.5. Then each agent updates its credence based on the results observed by himself and its neighbors. 
We start looking at the state of the network at its first instance \textit{t\textsubscript{1}}, at which we see a network of four nodes connected by a set of edges. At this instance, only one of the nodes votes for Beta action and the other three for Alpha action. It means that only one agent performs experiments. These experiments result in 7 successes in 10 tries, which supports the belief that Beta is better than Alpha action. Based on the gathered data, this agent and its neighbors update their levels of credence using Bayes' rule. One of the agents has no connections to anyone who performed experiments, so it has no reason to update its credence. At the state in \textit{t\textsubscript{2}} agents start the learning process with credence levels updated from previous learning iteration. For this state, one of the connections disappears and the network consists of just three active nodes. One of the nodes has no connection to anyone else at that time, so it is disabled and maintains its opinion.
At \textit{t\textsubscript{3}} we see two newly created edges and four active nodes. Evidence gathered through experiments still supports the Beta action superiority. In consequence, all agents change their credence and prefer Beta action. There is still no consensus, because the credence of all the nodes is not high enough -- over 0.99 or low enough -- less than 0.5.

\subsection{Dataset and Baseline Network Models}
\label{sec:dataset-and-baseline-models}
We run the simulations on both real social temporal networks built by applying the CogSNet model to the NetSense dataset, and static synthetic networks. Some of these static topologies have already been studied in previous work on learning in social networks~\cite{zollman2007communication}, while the others have been included due to their distinctive features. These synthetic networks are intended to serve as a reference for our temporal model. 
% Below we describe the NetSense dataset and static topologies that have been used in the experiments.

NetSense is an empirical dataset generated by human interactions~\cite{striegel2013lessons}. The data was collected among a group of about 180 students using a special application installed on their smartphones that recorded metadata on phone calls and text messages. The scope of the data covers three years, that is six semesters of studies of the group of students. In total, it contains 7,575,865 events, out of which 7,096,844 (93.7\%) are text messages and the rest are phone calls. Events were recorded for every student’s communication, including those with non-participants. Before processing the event sequence from the NetSense collection, we performed a series of data cleaning operations. To ensure data consistency, we decided to limit the events set only to activities between the study participants, as it was not possible to track activity between non-participants. In addition, we removed duplicates that were sometimes recorded by the data collection application. These operations limited the number of events to 537,575.

We utilized the NetSense dataset to create a temporal social network, which was later used as a topology for our temporal model. We applied the configuration of the CogSNet model introduced in~\cite{michalski2021social}, where it was validated on the same dataset. One day was a resolution of the generated temporal network. This means that as many as 1,103 points in time were generated, one for each day in which we run the decision process. We also model slower processes, in which the decisions are being made every one, three, five or ten days. The CogSNet model parameters are presented in Table~\ref{tab:cogsnet_params}.

\begin{table}
    \centering
    \caption{CogSNet parameters}
    \label{tab:cogsnet_params}
    \begin{tabular}{|l|r|}
    \hline
    \textbf{Parameter}  & \multicolumn{1}{l|}{\textbf{Value}} \\ \hline
    Forgetting function & \multicolumn{1}{l|}{exponential}    \\ \hline
    Trace lifetime      & \multicolumn{1}{l|}{3 days}         \\ \hline
    $\mu$                  & 0.3                                 \\ \hline
    $\theta$               & 0.2                                 \\ \hline
    $\lambda$            & 0.00563145983483561                 \\ \hline
    Unit                & \multicolumn{1}{l|}{1 hour}         \\ \hline
    \end{tabular}
\end{table}

The produced temporal network shows some seasonality in the activity of the participants, characterized by reduced communication during holiday breaks and inter-semester breaks, in agreement with an in-depth study of the seasonality presented in~\cite{kulisiewicz2018entropy}. In these periods the network is unstable, first a large number of nodes and connections disappear, and then reappear sometime later. In order to avoid the impact of such high instability on the study, we decided to limit the scope of experiments performed on the network to the period of the first semester, between 15 and 125 days. During this time, the number of active nodes and the number of edges is fluctuating, but the changes are not drastic, as in the case of the inter-semester breaks mentioned earlier. During the selected period, there were on average 148 nodes and 157 edges in the network.

\begin{figure}
    \centering\includegraphics[width=0.45\textwidth]{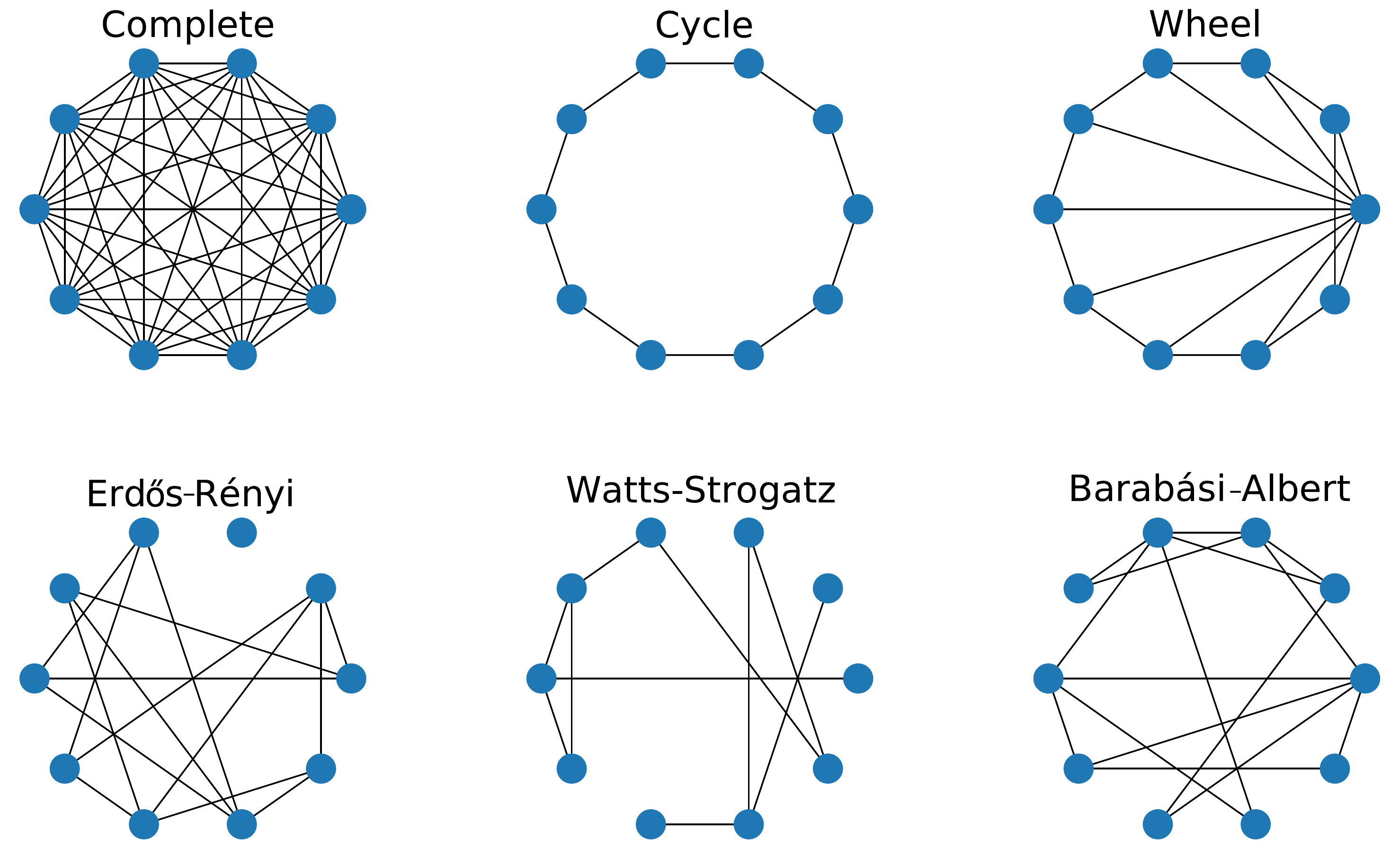}
    \caption{Example of six different static topologies, generated for a given size of 10 nodes.} 
    \label{fig:static-networks}
\end{figure}

In previous studies, the network epistemology model has been tested mainly for a relatively small number (a few to a few dozen nodes) of static topologies. For comparison, we also run simulations for other reference network topologies: complete, cycle, circle, and networks generated using the following models: Erdős–Rényi~\cite{erdos1960evolution}, Watts-Strogatz (WS)~\cite{watts1998collective}, Barabási–Albert (BA)~\cite{barabasi1999emergence} -- toy examples of these are depicted in Figure~\ref{fig:static-networks}.

The complete graph, cycle, and circle structures were studied in the previous work of Zollman\cite{zollman2007communication}, while the WS and BA networks show closer similarity to real social networks. The size and properties of the network can affect the network learning process. For this reason, we generated these reference networks in such a way as to provide similarity in size and properties to the studied temporal network. In the period under the study, the average number of active nodes in the temporal network was equal to 148.37 and the average number of edges between these nodes was equal to 157.38. According to that, the number of nodes was set to 148 for each reference network. In the Erdős-Rényi model, the edge occurrence probability was set to 0.0144. This value was calculated based on the fundamental property of the ER graph model that the expected number of edges $E$ the function of the probability $p$ of an edge between any pair of nodes, leading to equation $E = \binom{N}{2}p$, so $p = \frac{E}{\binom{N}{2}}$, where $N$ denotes the number of nodes. In the Watts-Strogatz model, the mean degree $K$ (base number of neighbors) was set to 2, which is the closest even integer (required by the model) to the value 2.12 calculated from the formula $K = \frac{2E}{N}$. The rewiring  (edge switching) probability was set to 0.5, which is a middle value from the available range \([0,1]\) and the default of the model. In the Barabási-Albert model, the number of edges to attach from a new node to existing nodes was set to 2. The choice of this value allows us to produce graphs with 293 edges, a number larger than desired, however this is the best approximation that could be obtained using BA model.

Another baseline network model was a snapshot of a temporal network that had been taken at the fifteenth day of students' interactions (shortly after the snapshots stabilized over time~\cite{kulisiewicz2018entropy}. We referred to it as \textit{static network} in this work, since there are no structural changes over time, and its structure and connectivity was closer to the temporal network than six aforementioned models.

\section{Results and Discussion}
In this section, we present the results of simulations performed with the proposed temporal model and the base static model for reference topologies. We use the NetSense dataset as a source for a temporal social network that has been built using the CogSNet model. Then, it is used in simulations of the learning process in the dynamic setting.

\begin{figure*}
\centering\includegraphics[width=\textwidth]{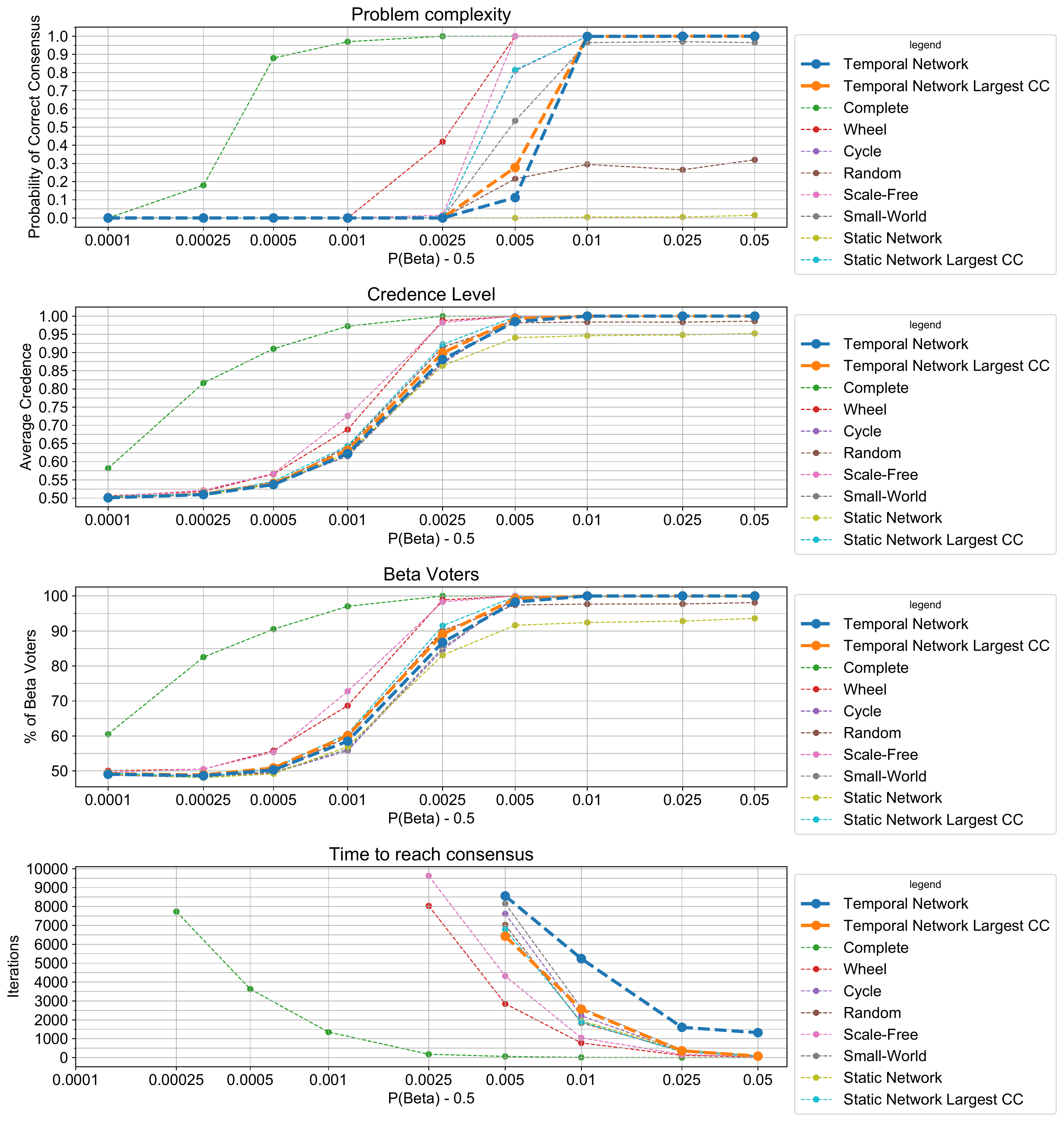}
\caption{Community performance for various cases of the given problem.} 
\label{fig:networkEffectiveness}
\end{figure*}

Figure~\ref{fig:networkEffectiveness} shows the effectiveness of a temporal network and reference networks in solving problems of various difficulties. The difficulty of the problem was controlled by changing the value of the Beta action payoff parameter. Other parameters of the model were predefined and constant. The \textit{Beta} action payoff values were taken from the set: \{0.5001, 0.50025, 0.5005, 0.501, 0.5025, 0.505, 0.51, 0.525, 0.55\}, that corresponds to the difference between the expected action payoffs: $\epsilon = P(Beta) - P(Alpha) = P(Beta) - 0.5$, from the set:  \{0.0001, 0.00025, 0.0005, 0.001, 0.0025, 0.005, 0.01, 0.025, 0.05\}. The maximum number of iterations for the static topologies was set to 10,000. To ensure a similar maximum number of possible iterations between the static models and the temporal network, the number of iterations per state of the temporal network was set to 91. This configuration resulted in the maximum number of iterations for the temporal network being set to 10,010, 10 more than for the static networks. Trials number \textit{N} was equal to 10. 1,000 simulations were run for each parameter configuration. It can be observed that all networks except Random and Small-World for Beta action payoff=0.501 reached the correct consensus for all simulations. For the two topologies mentioned above, this did not happen because some of the generated structures contained small connected components with just two nodes, which may have already received a drawn credence unsupportive of Beta action or have fallen into the trap of spurious data at the beginning of the simulation. This occurs when the credence of all individuals in a connected component falls below 0.5, so nobody is left to attempt to experiment with the Beta action and the entire connected component reaches incorrect consensus. Apart from the random and  Small-World graphs, for Beta action payoff values greater than 0.005, the probability of reaching correct consensus is lowest for both the entire temporal network and its largest connected component (CC) -- here, by the temporal network connected component we mean the component that has been established early in the temporal network and does not change significantly over iterations. Most importantly, however, the community in the temporal network configuration is not only able to reach consensus, although with a lower probability than in static networks, however for relatively simple problems all simulations ended up reaching the correct consensus on the action payoff. If we look at the average community credence, for simulations with beta action payoff equal to 0.505, we can see that it is comparable for all configurations. The lower likelihood of correct consensus for the temporal network may be caused by the presence of certain small connected components, separated from the rest of the network, falling into incorrect consensus, after which they had no opportunity to contact other nodes to change their minds. It is also worth noting that the average time to converge to consensus, regardless of the difficulty of the problem, was always the longest for the temporal network. Only this structure needed a relatively large number of nearly 1,500 iterations to converge to consensus, even for the easiest problem, where other networks were able to achieve it very quickly.

\begin{figure*}
    \centering\includegraphics[width=\textwidth]{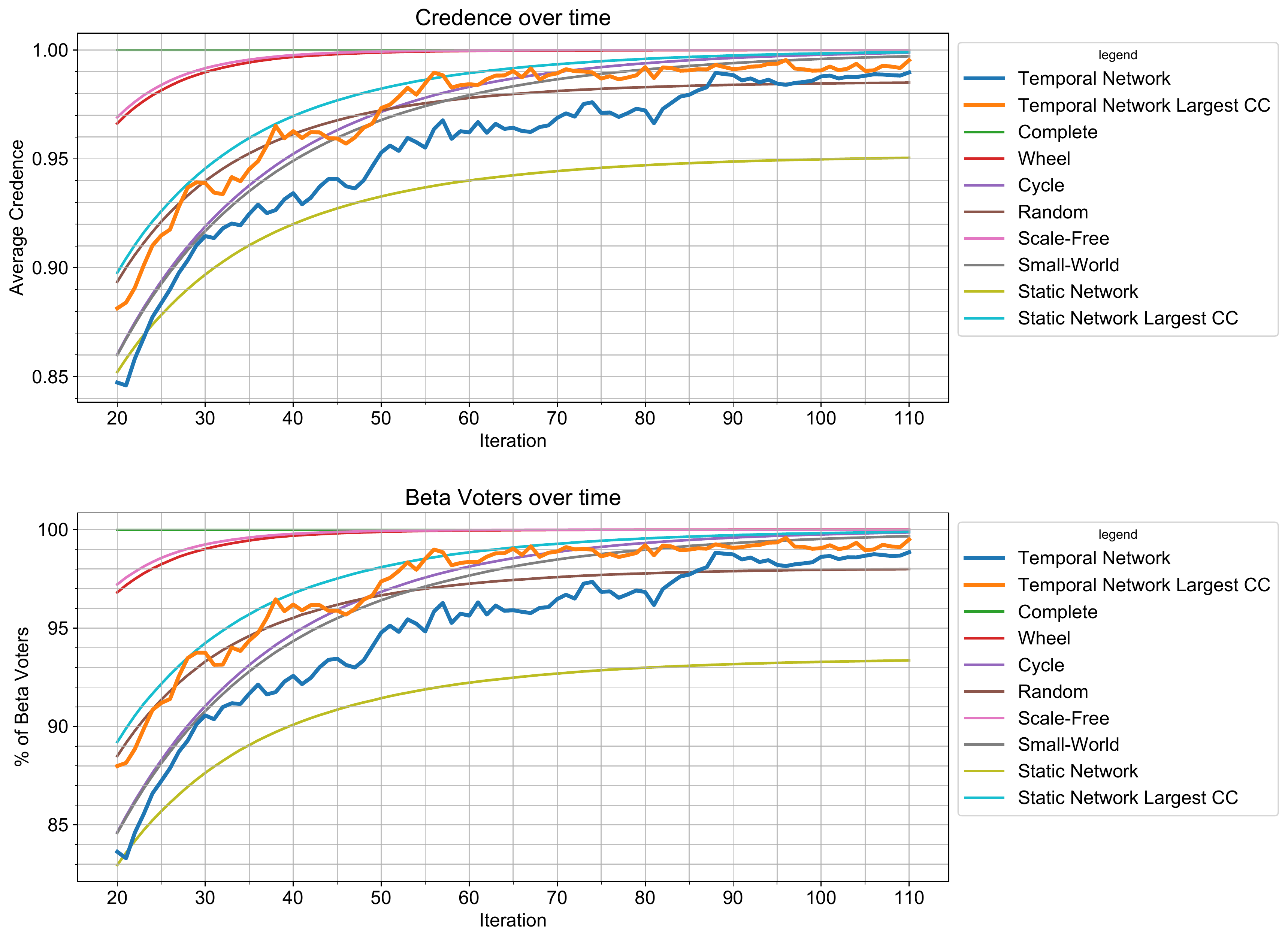}
    \caption{The course of the community learning process.} 
    \label{fig:learningProcess}
\end{figure*}

The first study shows the difference between results of the learning process between simulations with dynamic and static networks. Another interesting aspect is the behavior of the learning process itself. Figure~\ref{fig:learningProcess} shows the values of the average credence and the number of Beta action voters averaged over all simulations. A noticeable characteristic of the evolution of a temporal network over time is its instability and fluctuating size of its largest connected component (CC). The average values of both measured characteristics, for the static networks, seem to be non-decreasing over time, and converging in the case of credence to 1 and in the case of beta voter fraction to 100 percent. Instability arises only in the temporal network, even when the results are averaged over a large number of simulations. This observation indicates that the dynamic structure of the network causes fluctuations in learning. These fluctuations may be a reflection of new nodes joining the network or reactivating nodes that fell out of the structure some time earlier and were inactive for some time. The chance that such nodes have a low level of credence and choose the Alpha action is greater than that they are followers of the Beta action, especially if they were previously elements of smaller coherent components which are easier to fall into a state of incorrect consensus. The research presented in the introduction shows that network structure is an important factor in the process of opinion formation in social networks. Here examine how dynamic structure affects the time evolution of this process. Communities organized in reference network structures, and in a dynamic network were given a relatively simple problem to solve. In this case, the goal is not so much to see how successfully each community will perform, but the process itself. To ensure comparisons, all communities performed an identical number of iterations, that is, 110. In the dynamic network, one iteration was performed for one time window -- it is as if individuals performed experiments and contacted each other to update their views exactly once a day for 110 days. In this study, 10,000 simulations were run for each network. Trials number \textit{n} was equal to 10.

The first important observation is that, at least for the temporal network based on the NetSense dataset, the community despite its dynamic structure can reach the correct consensus. The probabilities of reaching the correct consensus are different for different levels of problem difficulty and decrease as the problem gets harder. We can also observe that the chance of convergence to the correct consensus by the temporal network is lower than for most of the reference network models.

The lower performance of the Random networks is caused by the choice of parameters used to generate them. For the used parameters, the Erdos-Renyi graph results with high probability in the creation of a network having more than one connected component. These connected components may consist of a small number of nodes whose randomly initiated beliefs lead them to choose the Alpha action and not to experiment. This configuration and the static structure of the community do not allow the members of these connected components to change their beliefs, thus blocking the whole network from reaching a correct consensus.

Single cases of convergence to incorrect consensus occurred only for the complete network, and for the more difficult problems. This phenomenon of dense connectivity leading to a negative influence on the learning process was described as the Zollman effect \cite{zollman2007communication}. The fact that we don't observe any cases of convergence to the incorrect consensus for other networks, is due to the size of the networks. This relationship of convergence probability with network size was described in \cite{oconnor_rosenstock_2017}. 

For the static network setting, in which the network is a single snapshot of a CogSNet model, the joint belief formation 
is highly unlikely. Since there is no opportunity to share knowledge between individuals located in different connected components, the only scenario is that each component starts with the same initial opinion. The situation is different for the largest connected component of a static network that was separately analyzed as well. Here, the consensus had can be reached rather fast. Unlike convergence for a temporal network seen in Figure~\ref{fig:learningProcess}, the learning process here is smoother as there are no network fluctuations. Another perspective on the results arises from the study of entropy evolution of social interactions and communication~\cite{michalski_2020, kulisiewicz2018entropy}. From this perspective, the high probability of reaching consensus in the temporal setting is caused by the contribution of the dynamics itself. Over time, the opinions can be carried around by people connecting to different parts of the network. Thus, the limitations of clustering can be overcome in way similar to what Kurahashi-Nakamura et al. observed in opinion dynamics~\cite{kurahashi2016robust}, where connecting agents with smaller disagreement eventually led to reaching consensus.

\begin{figure*}
    \centering\includegraphics[width=\textwidth]{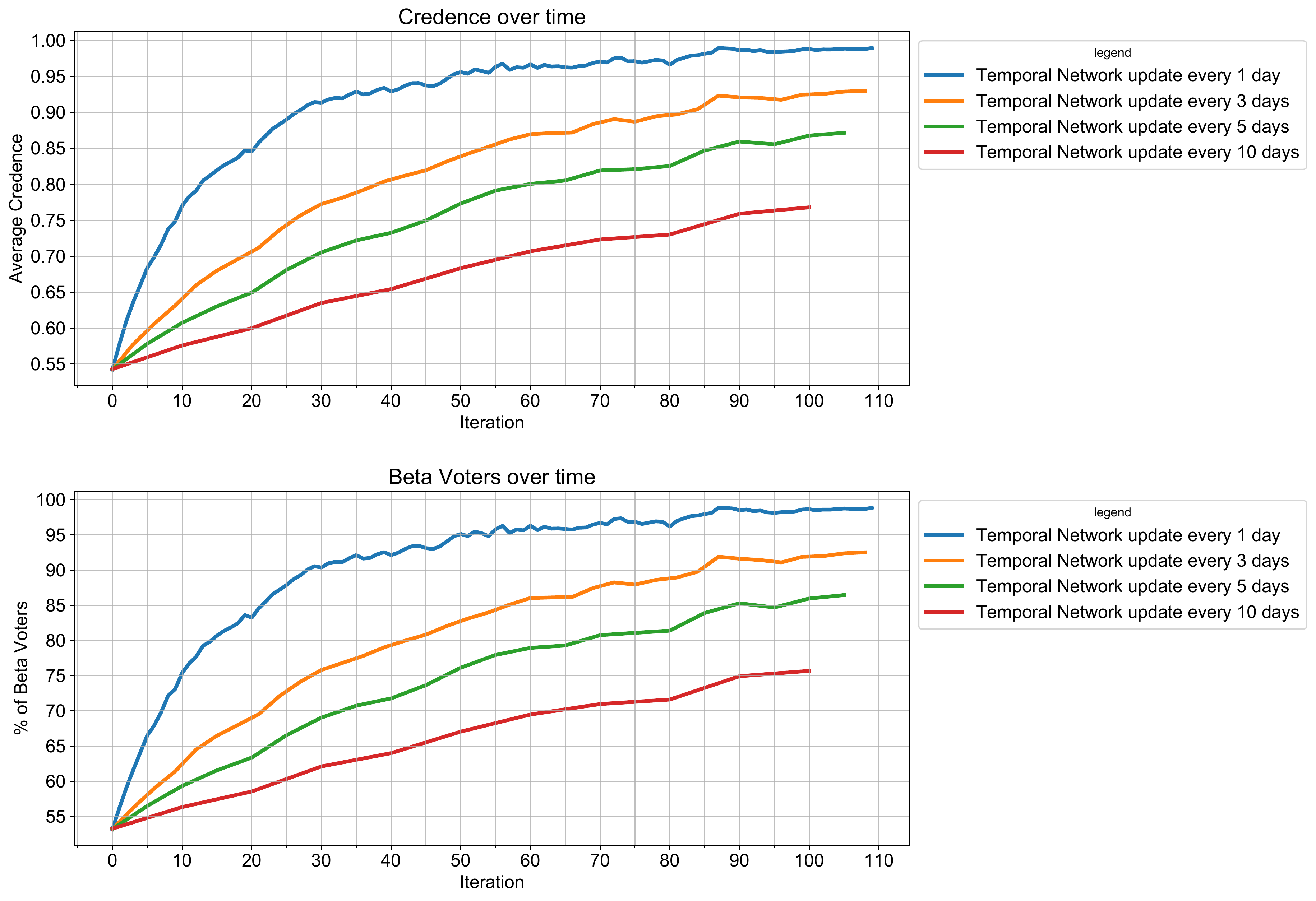}
    \caption{The analysis of the learning process with different timescales: updating the state of agents every day, or three, five, or ten days.}
    \label{fig:credenceDifferentScale}
\end{figure*}

In another set of experiments, we investigated how the learning process evolves in a situation when its iterations are not aligned with the temporal network. As described in Section~\ref{sec:dataset-and-baseline-models}, we conducted the experiments for updating the actors' states every one, three, five and ten days. The rationale for slightly longer periods for updating credence was that people may require different periods of time to digest an idea before intrinsically accepting or rejecting it. Given the experiments were conducted in the university setting, we selected these periods on the scale from one to ten days to be in line with the best recall performance found in~\cite{michalski2021social}. The results demonstrate (see Figure~\ref{fig:credenceDifferentScale}), that the trend is the same - the percentage of beta voters increases over iterations, as well as the average credence. However, the dynamics of the process is slower. The conclusion is that despite the fluctuations introduced by network dynamics, the process follows the same direction, only the speed is affected.

Now, we look at the parameter space of the learning process. The experiment confirms that the network epistemology model is sensitive to the choice of parameters. A small difference in Beta action payoff can determine the ability of a community to collectively draw correct conclusions and come to a consensus on a topic in a finite time.

If we look at the results for the Beta payoff equal to 0.505, we can see that for the temporal network the probability of reaching the correct consensus for the largest connected component of the temporal network is higher than for the whole network. Even though the whole community may be struggling to solve the problem posed to it, the largest subgroup of the network reaches a correct consensus. This means that those who rejected better action tend to be on the periphery of the community, gathered in smaller groups supporting the same, but worse, view. The same applies to the number of iterations needed to converge to a consensus. In the full network, it is significantly longer than for static reference networks and for the largest connected component. Convincing initially isolated individuals requires contacting them and cannot be done any other way. Isolated individuals who insist on an erroneous view may and will change their minds, but they must be given evidence to do so.

The dynamic structure of the community also affects the process of learning itself. The number of people advocating different options can change dynamically and affect the proportion of groups supporting different actions. The average credence and the number of agents supporting the better action don't decrease for any of the reference network structures as it does for the temporal network.

It is important to note, however, that the study was conducted on only a fragment of one temporal network and requires confirmation from the large studies that we left for future research. Our goal was to support the predictions about the effect of dynamized structure on learning and we believe that our results accomplished that.

\section{Conclusions}
This work introduces a temporal network epistemology model that corresponds to reality closer than static epistemic networks studied in previous studies on knowledge exchange and decision making. By applying the CogSNet model we were able to build a temporal social network through which agents gather information to decide their next actions.

As the results demonstrate, accounting for dynamic topology reveals the existence of a number of effects that either prohibit or significantly slow down reaching the consensus compared to static social structures. This is naturally caused by network dynamics, but two major aspects of it contributed the most, namely nodes' isolation and varying neighborhood of nodes. This clearly shows that representing real-life phenomena in the network dynamics reveals significant divergence from the simplified outcomes of typical epistemological studies based on static networks. Moreover, contrary to the models based on static networks, it should be also expected that different connected components will end up with different states regarding the consensus -- the longer this situation persists, the harder it would be to direct the network towards the expected situation if some links between these components reappear in the future.

Given that the goal of this study was to analyze the process of the decision-making process in temporal epistemic networks, a variety of new insights have been drawn, but also a number of future work directions have also been found interesting. Among the latter, we believe that it would be beneficial to investigate the outcomes of the process for different sets of CogSNet model parameters. So far, we only studied the most plausible values as past research on forgetting has shown~\cite{michalski2021social}. However, for certain processes, these values can be different and an interesting research direction would be to answer the question of their impact on the process. Next, given that real-world social networks do consist of separate components, we study how different configurations and their dynamic interconnections shape this process, especially that we can see that the number and structure of connected components significantly impact the results. For instance, does a series of one-day links that appear every second day contribute more than a link that exists continuously but only for the first half of the analyzed period? Next, given that the credence update period can mean something different for different individuals, as our interactions highly vary, another idea would be to update the credence based not on a time passed, but rather a number of interactions made with others. We plan to study these ideas in the future analysis of temporal epistemic networks in the context of multiple dependencies that network dynamics introduces.

\section*{Data Availability}
The code for computing the state of the social network built using the CogSNet model has been made public. A fast C implementation is available as a Code Ocean capsule\footnote{\href{https://dx.doi.org/10.24433/CO.6088785.v1}{https://dx.doi.org/10.24433/CO.6088785.v1}}. Access to the NetSense dataset for research use can be gained by contacting Professor Omar Lizardo from University of California Los Angeles.

\section*{Funding Statement}
BKS was partially supported by DARPA Social-Sim Program No. W911NF-17-C-0099, and INCAS Program No. HR001121C0165. PK was partially supported by the National Science Centre, Poland, project no. 2016/21/B/ST6/01463. RM and MN were partially supported by the National Science Centre, Poland, project no. 2021/41/B/HS6/02798.

\section*{Author contributions}
All authors participated in the design of the research and computational experiments. DS ran the computational experiments and collected the results. All authors participated in analysis of the results. All authors wrote and edited the manuscript.

\section*{Conflicts of Interest}
Authors declare no conflict of interest.

\bibliographystyle{plain}
\bibliography{bibliography}

\end{document}